\documentclass[twocolumn,pra,aps]{revtex4}
\usepackage{graphicx}
\usepackage{amsmath}

\begin{document}
\title{Long distance quantum teleportation of qubits from photons at 1.3\,$\mu m$ to photons at 1.55\,$\mu m$ wavelength}
\author{I. Marcikic$^\dag$, H. de Riedmatten$^\dag$, W. Tittel$^{\dag,\ddag}$, H. Zbinden$^\dag$ and N.
Gisin$^\dag$}

\affiliation{$^\dag$Group of Applied Physics-Optique, University of
Geneva, CH-1211, Geneva 4,
Switzerland\\
$^\ddag$QUANTOP - Danish National Research Foundation Center for
Quantum Optics, Institute for Physics and Astronomy, University of
Aarhus, Denmark}

\begin{abstract}
Elementary 2-dimensional quantum states (qubits) encoded in
1.3\,$\mu m$ wavelength photons are teleported onto 1.55\,$\mu m$
photons. The use of telecommunication wavelengths enables to take
advantage of standard optical fibre and permits to teleport from
one lab to a distant one, 55\,m away, connected by 2\,km of fibre.
A teleportation fidelity of 81.2\,\% is reported. This is large
enough to demonstrate the principles of quantum teleportation, in
particular that entanglement is exploited. This experiment
constitutes a first step towards a quantum repeater.
\end{abstract}

\maketitle

\section{Introduction}

According to Aristotle, objects are constituted of {\it matter}
and {\it form}. Today one would say energy and structure. For
quantum physicists objects are constituted by {\it elementary
particles} and {\it quantum states}. Matter and energy can not be
teleported: they can not be transferred from one place to another
without passing through intermediate locations. However, in 1993
it was discovered \cite{brassard93} that quantum states (i.e. the
ultimate structure of objects) can be teleported: this is quantum
teleportation. Accordingly, objects can be transferred from one
place to another without ever existing anywhere in between! But
only the structure is teleported, the matter stays at the source
side and has to be already present at the final location.
Moreover, the matter at the final location has to be entangled
(i.e. form an EPR state \cite{einstein35}) with yet another piece
of matter, located near the original object. The original object
and the nearby half EPR state undergo then a joint measurement,
i.e. a so-called {\it Bell state measurement}, whose result is
communicated to the distant final location where it determines a
simple transformation to be applied to the second half of the
EPR-state. This second half carries now precisely the quantum
state of the original object. Note that the Bell state measurement
destroys the quantum state of the initial object and that no
information about which state is teleported is acquired because
its final state is completely mixed. The first, and with
foreseeable technologies the only, application of quantum
teleportation is in quantum communication where it helps to extend
quantum cryptography to larger distances \cite{gisin02}.

Since the famous article by Bennett and five colleagues presenting
the concept \cite{brassard93}, quantum teleportation received a
lot of attention. On the conceptual side, it has been proven that
it is a universal gate for quantum computing \cite{gottesman99}.
In particular, together with quantum memories, it offers the
possibility to realize quantum repeaters with unlimited range
\cite{briegel98}. But it is fair to say that its fundamental
meaning for our understanding of quantum nonlocality and of the
structure of space and time may still awaiting discovery. On the
experimental side, progress in demonstrations of the concept has
been surprisingly fast
\cite{boschi98,bouwmeester97,furusawa98,nielsen98,kim01,lombardi02,babichev02,bowen02,jennewein02}.
Already in 1997, only 4 years after Bennett et al.'s landmark
publication, two groups, one in Rome, one in Innsbruck, presented
first results of quantum teleportation employing qubits (i.e. two
dimensional quantum states). The Italian group, led by Prof. de
Martini \cite{boschi98}, exploited an idea by Prof. Popescu to
teleport a qubit carried by one of the photons of the EPR pair.
However, this nice trick prevents the possibility to concatenate
this teleportation scheme. The Austrian group, led by Prof.
Zeilinger \cite{bouwmeester97} (now in Vienna), used a more
complete scheme, where only one qubit is carried per photon.
Hence, soon after their initial experiment they could also
demonstrate entanglement swapping \cite{zukowski93,pan98}, i.e.
the teleportation of an entangled qubit. However, this scheme was
also incomplete since it used what is called {\it a partial Bell
state measurement} which implies that even in principle the
teleportation succeeds only in 25\,\% of cases. A few years later,
Prof. Shih and his group \cite{kim01} at Baltimore University
reported on a teleportation experiment with complete Bell state
measurements, but the efficiency of the measurement was only of
the order $10^{-10}$. The difficulty is, as proven by L\"utkenhaus
et al. \cite{lutkenhaus99}, that a complete Bell state measurement
for qubits requires non-linear optics. Hence, either one stays
with linear optics and admits incomplete measurements (as
Zeilinger's group and the result presented in this article); or
one goes for non-linear optics and admits very inefficient
measurements; or one does not use qubits, nor finite dimensional
Hilbert spaces. Indeed, it has been shown, first theoretically
\cite{braunstein98}, then experimentally \cite{furusawa98}, that
the teleportation of continuous variables can, in principle, be
fully achieved using linear optics. However, the difficulty is
then to produce close to maximally entangled EPR states. This
difficulty is even more significant when the distance is
increased, because squeezed states of light beams (i.e. EPR states
for continuous variables) are very vulnerable to losses.

In this article we report the first experimental long distance
demonstration of this fascinating aspect of quantum mechanics.
Qubits carried by photons of 1.3\,$\mu m$ wavelength are
teleported onto photons of 1.55\,$\mu m$ wavelength from one lab
to another one, 55\,m away in bee line and connected by 2\,km of
standard telecommunication fibre. Our experiment follows the line
of Zeilinger's group in that we use linear optics for our partial
Bell state measurement. However, it also differs significantly in
that our qubits and necessary EPR-state are not encoded in
polarization, but in superposition and entanglement of time-bins
\cite{tittel01,brendel99,thew02}, respectively. Moreover, we use
two nonlinear crystals, which is necessary for the implementation
of quantum communication protocols where space like separation of
the photon pair sources is required \cite{briegel98}. Each of them
produces, by spontaneous parametric downconversion, pairs of non
degenerate photons at telecommunication wavelengths of 1.3 and
1.55\,$\mu m$ which enables us to employ standard optical fibres,
and to transmit the photons over large distances.

\section{Time-bin qubit}

Qubits can be realized in an unlimited number of ways. A
well-known one uses photon polarization, but this is not the
optimal choice for long distance quantum communication. Another
implementation of qubits in photons, more robust against
unavoidable polarization fluctuations in fibres, consists in using
time-bins \cite{tittel01,brendel99,thew02}. Figure 1b presents the
preparation of an arbitrary qubit state as superposition of two
time-bins $c_0 \left|1,0 \right\rangle+c_1e^{i\phi}\left|0,1
\right\rangle$. This state corresponds to the input photon passing
through the short arm $\left|1,0\right \rangle$ of an unbalanced
interferometer, with probability amplitude $c_0$, and through the
long arm $\left|0,1\right \rangle$, with probability amplitude
$c_1$. The phase $\phi$ characterizes the imbalancement of the
interferometer with respect to a reference optical path length
difference. Alice's variable coupler allows one to choose the
value of the two amplitudes, and the switch enables to superpose
them without losses.

\begin{figure}[h]
\includegraphics[width=8.43cm]{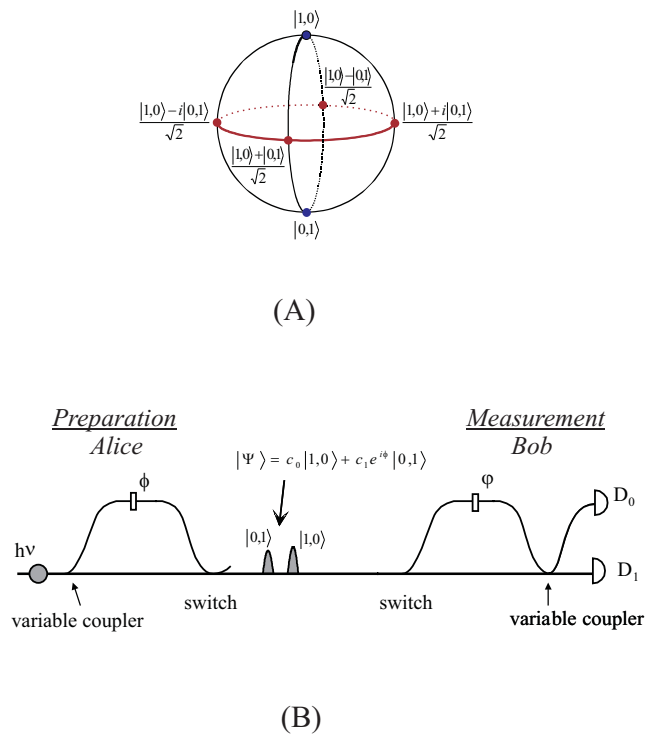}
\caption{{Principle of preparation and measurement of time-bin
qubits. \textbf{(A)} Poincaré (qubit) sphere. The states
$\left|1,0\right \rangle$, $\left|0,1\right \rangle$ and their
superpositions $\left|1,0 \right\rangle+e^{i\phi}\left|0,1
\right\rangle$ are represented on the north pole, south pole, and
on the equator, respectively. \textbf{(B)} Schematic
representation of the preparation and the measurement of time-bin
qubits. Using a variable coupler Alice chooses the probability
amplitudes of the first and the second time-bin (latitude on the
qubit sphere), and the phase shifter defines the relative phase
$\phi$ (the longitude on the qubit sphere). Bob uses an similar
device to choose the measurement basis. }} \label{time-bin}
\end{figure}

Figure 1 also shows how arbitrary (projective) measurements can be
implemented. Bob's switch is used to send the first time-bin
through the long arm, and the second time-bin through the short
arm such that they arrive simultaneously at the beam-splitter.
With the phase shifter and the variable coupler we can choose to
measure the state in any basis \cite{tittel01}.

In the experiment, instead of using a true variable coupler, we
use 3 different settings with coupling ratio of 0\,\%, 100\,\% and
50\,\%. These settings correspond to preparation of and
projections onto the states represented on the north pole, south
pole and on the equator of the generalized Poincaré sphere (see
\ref{time-bin}). We did also replace the switches by passive fibre
couplers. This implies a 50\,\% loss, both for the preparation and
the measurement apparatuses. But since fast switches have even
larger losses, our choice is the most practical one and does not
affect the principle of the experiment. The result of the
measurement for each basis can then be found by looking at the
appropriate detection times \cite{tittel00}. Note that the concept
of time-bins, contrary to polarization, can easily be generalized
to higher dimensions \cite{riedmatten02}. Time-bins are sensitive
to chromatic dispersion, but this phenomenon can be passively
compensated using linear optics \cite{zbinden01}, on the contrary
to polarization mode dispersion \cite{noe99}.

\section{Experimental realization}

\begin{figure}[h]
\includegraphics[width=8.43cm]{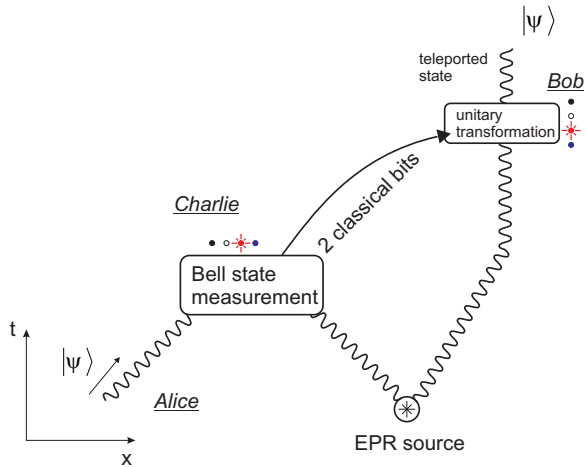}
\caption{{Quantum teleportation protocol: Alice sends an unknown
qubit $\left|\Psi \right \rangle$ to Charlie. Charlie shares with
Bob a pair of entangled qubits emitted by the EPR source, and can
communicate with him. Charlie now performs a joint measurement on
Alice's qubit and his part of the entangled pair, thereby
projecting the two qubits state onto one of the four Bell states.
Depending on the result, Bob performs a unitary transformation to
recover the initial state $\left|\Psi \right \rangle$ of Alice's
qubit.
 }} \label{qtscheme}
\end{figure}

Let us recall the quantum teleportation scenario and define the
terminology. Suppose that Alice wants to transmit an unknown
quantum state of her qubit to Bob. However, she cannot send the
particle directly, for instance due to a lossy transmission
channel. But she has the possibility to send her qubit to Charlie
who shares a pair of entangled qubits and a classical
communication channel with Bob (see fig.\ref{qtscheme}). Charlie
now entangles Alice's qubit with his part of the shared pair by
means of a so-called Bell state measurement \cite{michler96} and
then communicates the result, i.e. the Bell state he projected
onto, to Bob. Performing now an unitary transformation - a bit
flip, a phase flip, or both, or the identity operation -,
depending on Charlie's result, Bob's photon ends up in the initial
state of Alice's photon, although the state remains unknown.

\begin{figure}[h]
\includegraphics[width=8.43cm]{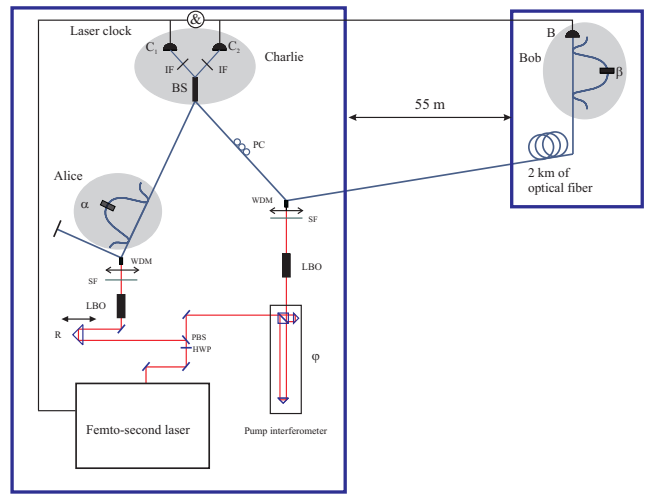}
\caption{{Experimental setup: Femtosecond laser pulses are split
in two parts using a variable beam-splitter (HWP+PBS). The
reflected beam is sent to Alice who creates the time-bin qubits to
be teleported (at 1310\,nm), which she then forwards to Charlie.
The transmitted beam is used to produce non-degenerate entangled
time-bin qubits. The 1310 nm photon is sent to Charlie, the
1550\,nm photon to Bob who is situated in another lab, 55\,m away
from Charlie and connected by 2\,km of optical fibre. Charlie
performs a partial Bell state measurement on Alice's qubit and his
part of the entangled pair, using the 50/50 fibre coupler BS.
Whenever the two particle state is projected onto the
$\left|\Psi^{-} \right \rangle$  Bell state, Bob analyses his
photon to prove that indeed the state encoded by Alice has been
teleported.}} \label{qtsetup}
\end{figure}

Our experimental set-up is presented in fig.\ref{qtsetup}. A mode
locked Ti:Sapphire laser (Coherent Mira 900) produces 150\,fs
pulses at $\lambda_p=710\,nm$ with a repetition rate of 76\,MHz.
To remove all unwanted infrared light the beam passes through a
series of dichroic mirrors, reflecting only wavelengths centred
around 710\,nm. The laser beam is then split into two parts using
a variable beam-splitter made of a half wave plate (HWP) and a
polarizing beam splitter (PBS).

The transmitted (horizontally polarized) beam is used to create
entangled time-bin qubits (EPR source) by passing the beam first
through an unbalanced bulk Michelson interferometer (referred to
as the pump interferometer) with optical path-length difference
$\Delta \tau=1.2\,ns$. The beam is then directed onto a type I
non-linear crystal (Lithium triborate, LBO, from Crystal Laser)
where entangled non-degenerate collinear time-bin qubits at
telecom wavelengths (1310 and 1550\,nm) are created. The pump
light is removed with a silicon filter (SF) and the twin photons
are collimated into an optical fibre and separated by a
wavelength-division-multiplexer (WDM). The 1310\,nm photon is then
sent to Charlie and its twin 1550\,nm photon to Bob. The entangled
state is described by:

\begin{equation}
\left| \Phi \right\rangle =\frac{1}{\sqrt{2}}(\left|
1,0\right\rangle _{Charlie}\left| 1,0\right\rangle
_{Bob}+e^{i\varphi }\left| 0,1\right\rangle _{Charlie}\left|
0,1\right\rangle _{Bob}) \label{1}
\end{equation}
where $\left|1,0\right \rangle$ represents the first time-bin and
$\left|0,1\right \rangle$ the second one. The imbalancement of
this interferometer defines the reference time difference between
the first and the second time-bin, thus the phase $\varphi$ is
taken to be zero. As described in \cite{marcikic02} we observed
two-photon fringe visibilities up to 95\,\% when testing the
source, showing that the purity and the degree of entanglement is
high enough for the use in quantum communication protocols.

The reflected (vertically polarized) beam is used to create the
qubits to be teleported. Similar to the creation of the entangled
pairs, the beam is focussed into a LBO crystal creating pairs of
photons at 1310\,nm and 1550\,nm wavelengths, however, without
passing first through an interferometer. After blocking the pump
light and coupling the photon pairs into an optical fibre, the
1550\,nm photon is removed using a WDM. Alice passes the 1310\,nm
photon through an unbalanced fibre Michelson interferometer,
thereby creating a time bin qubit:

\begin{equation}
\left| \Phi \right\rangle_{Alice} =\frac{1}{\sqrt{2}}(a_0\left|
1,0\right\rangle _{Alice}+a_1e^{i\alpha}\left| 0,1\right\rangle
_{Alice}) \label{2}
\end{equation}
where 0,1 or $\frac{1}{\sqrt{2}}$ depending on the discrete
variable coupler setting, and $a_1=\sqrt{1-a_0^2}$. The phase
$\alpha$ is defined relatively to the reference phase $\varphi$.
Alice's qubit   is finally sent to Charlie. Charlie performs the
joint Bell-state measurement between the qubit sent by Alice and
his part of the pair, with the 50/50 fibre coupler BS. As proven
by L\"utkenhaus et al. \cite{lutkenhaus99}, only two out of four
different results can be discriminated using linear optics. We
choose to select only the one which projects the two particles
onto the singlet entangled state

\begin{equation}
\left| \Phi^- \right\rangle =\frac{1}{\sqrt{2}}(\left|
1,0\right\rangle _{Alice}\left| 0,1\right\rangle _{Charlie}-\left|
0,1\right\rangle _{Alice}\left| 1,0\right\rangle _{Charlie})
\label{3}
\end{equation}
This takes place when the two photons trigger the detectors
labelled $C_1$ and $C_2$ on fig.\ref{qtsetup} at times that differ
precisely by the time difference $\Delta \tau$ between two
time-bins. Indeed, each of the two terms in Eq.\ref{3} may produce
this detection result, either with each photon remaining in their
fibre or both coupling to the other (hence the $\pi$ phase shift
which corresponds to the minus sign in Eq.\ref{3}). To achieve
this projection the two photons have to be indistinguishable when
they emerge from the beam-splitter \cite{riedmattendip02}, i.e.:
(i) They have to be localised in the same two spatial output
modes. This condition is easily met using a single mode fibre
beam-splitter. (ii) The two photons must arrive at the beam
splitter at the same time, within their coherence time $\tau_c$.
This means that the pump pulses have to be better localised than
the created photons. This condition is fulfilled by using 150\,fs
pump pulses together with 10\,nm filters (IF) centred at 1310\,nm
to increase the coherence time of the created photons to about
$\tau_c=250\,fs$. The arrival time of Alice's qubit is controlled
by a retroreflector (R) mounted on a micrometric translation
stage. (iii) The time separation $\Delta \tau$ between two time
bins for Alice's qubit and the entangled qubit must be the same,
again within $\tau_c$. Thus, the optical path length difference of
the pump and Alice's interferometer have to be precisely aligned.
This is achieved using white light interferometry and then, for
the fine alignment, by maximizing the visibility of the two photon
interference \cite{marcikic02}. (iv) The spectra of the two
photons must be equal. This is insured by using the same PDC
process and by the IF. (v) The polarisation of the two photons
have to be the same when they arrive at the beam-splitter. This is
done with the polarisation controller (PC), aligned with a LED and
a classical polarimeter. Photon $C_1$ is detected by a passively
quenched Germanium avalanche photo diode (APD) working at liquid
nitrogen temperature in so called Geiger mode
\cite{owens94,ribordy98}(quantum efficiency $\eta =10\,\%$, dark
count rate dc=35\,KHz, from NEC). To reduce the noise we make a
coincidence between the Germanium APD and a trigger from the laser
pulses. Photon $C_2$ is detected with a Peltier cooled InGaAs APD
working in the so called Gated mode \cite{stucki01} ($\eta =
30\,\%$, dc=$ 10^{-4}$ per ns, from id Quantique). This means that
the APD is only active during a short time gate ($\sim 100\,ns$)
when a photon is expected. The trigger is given by the coincidence
between the Germanium APD and the laser pulse. Fast electronics
provides information about the arrival time with an accuracy of
600\,ps. Finally the signals of the APDs are sent to a fast
coincidence electronics to achieve the Bell state measurement.

As we reported in \cite{marcikic02} the production of multi-photon
pairs by the EPR source should be avoided in many quantum
communication protocols. If the probability of creating a photon
pair is the same in both crystals, then, due to stimulated
emission, the probability of creating two pairs in one crystal is
the same as the probability of creating one pair in each crystal
\cite{linares01}. Thus, two times over three Charlie detects a
wrong event \cite{riedmattendip02}. In order to detect only the
desired events, we decrease the probability of creating entangled
qubits relative to the probability of creating the qubit to be
teleported. The wrong events are thus reduced to only the cases
where two entangled pairs are created, which can be made arbitrary
small. The ratio of probabilities is controlled by the variable
coupler HWP+BPS. Eventually we chose the ratio of 8 with the
probability of creating Alice's qubit per laser pulse of around
10\,\%. Bob is situated in another lab, 55 m away from Charlie in
bee line. To simulate a longer distance we added 2 km of
dispersion shifted optical fibre before the teleported photon
reaches Bob's analyser. Once Charlie has the information that the
(partial) Bell state measurement was successful, he informs Bob by
the classical channel. This operation projects Bob's photon onto
the state:

\begin{equation}
\left| \Phi \right\rangle_{Bob} =\frac{1}{\sqrt{2}}(a_1e^{i\alpha
}\left| 1,0\right\rangle _{Bob}-a_0\left| 0,1\right\rangle _{Bob})
\label{4}
\end{equation}

In order to recover Alice's qubit state (Eq.\ref{2}) Bob should
apply the $\sigma_y$ unitary transformation consisting in a bit
flip ($\left| 1,0\right\rangle \leftrightarrow \left|
0,1\right\rangle$ ) and a phase flip (a relative phase $\pi$).
However these unitary operations are not necessary to prove that
teleportation takes place. To show that our teleportation set-up
operates correctly, Bob analyses the received photon with an
analyser adapted for the wavelength of 1550\,nm \cite{marcikic02}
(see Fig.\ref{time-bin}). The analysis basis thus contains the
vector:

\begin{equation}
k_0\left| 1,0\right\rangle+k_1e^{i\beta}\left| 0,1\right\rangle)
\label{5}
\end{equation}
where 0, 1 or $\frac{1}{\sqrt{2}}$, depending on the variable
coupler setting, and $k_1=\sqrt{1-k_0^2}$.

Bob's photon is also detected with a InGaAs APD, as photon $C_2$.
The trigger is again given by the coincidence between the
Germanium APD and the laser pulse, but delayed by ~10 $\mu s$
which correspond to the time that the photon needs to propagate
down the 2\,km of fibre and arrive at Bob's detector. The
detection of Bob's photon triggers a 1310\,nm laser pulse. This
pulse is sent back to the primary lab through another fibre and
detected with a standard PIN diode. This detection is reported to
the coincidence electronics as the detection of photon $B$.
Finally we monitor four-fold coincidences with a time to amplitude
converter (TAC), where the start is given by a successful Bell
state measurement (detectors $C_1$ and $C_2$ + laser pulse) and
the stop by detector $B$. The start plays the role of the
classical information that Charlie sends to Bob. We choose to
record only the events when the photon in output $C_2$ arrives
with a time difference $\Delta \tau$ after the photon in output
$C_1$. We also record the coincidence between detectors $C_1$ and
$B$. The rate should remain constant, as it contains no
information about the Bell state measurement result. This provides
a control of the stability of the entire set-up.

\section{Results}

In order to show that our teleportation set-up is universal we
report the teleportation of two different classes of states. The
first one is composed of superpositions of two time-bins, hence
corresponds to states represented by points on the equator of the
Poincar\'e sphere (see Fig.\ref{time-bin}). The second class
consists of the two time-bins themselves, represented by the north
and south poles of the sphere. The quality of the teleportation is
usually reported in terms of fidelity $\bar{F}$ , i.e. the
probability that Bob's qubit successfully passes an analyzer
testing that it is indeed in the state $\Psi_{Alice}$ prepared by
Alice, averaged over all possible $\Psi_{Alice}$:

\begin{equation}
\bar{F} =\int{\left \langle \Psi_{Alice}|
\rho_{out}|\Psi_{Alice}\right\rangle d\Psi_{Alice}} \label{6}
\end{equation}
The linearity of quantum mechanics implies

\begin{equation}
\bar{F} =\frac{2}{3}F_{equator}+\frac{1}{3}F_{poles} \label{7}
\end{equation}
where $F_{equator}$ and $F_{poles}$ are the averaged fidelities
for the equatorial and pole states, respectively. To measure the
teleportation fidelity $F_{equator}$ of the equatorial states we
scanned the phase $\beta$ in Bob's interferometer (with both
discrete variable couplers at setting 50\,\%). This results in the
normalized coincidence count rate:

\begin{equation}
R_{C}=\frac{1-Vcos(\alpha+\beta)}{2}
\label{8}
\end{equation}
corresponding to Bob's state
$\rho_{out}=V\left|\Psi_{Alice}\right\rangle\left\langle\Psi_{Alice}\right|+(1-V)\frac{1}{2}$
, where V is the visibility which can theoretically reach the
value of 1. Accordingly, the fidelity equals 1 with probability V
and equals $\frac{1}{2}$ with probability 1-V, hence
$F_{equator}=\frac{1+V}{2}$. Figure 4a shows a nice result leading
to a fidelity of $(85\pm2.5)$\,\%. By performing repeatedly many
experiments over a few weeks with different phases $\alpha$ we
typically obtain fidelities around $(80.5 \pm 2.5)\,\%$.

\begin{figure}[h]
\includegraphics[width=8.43cm]{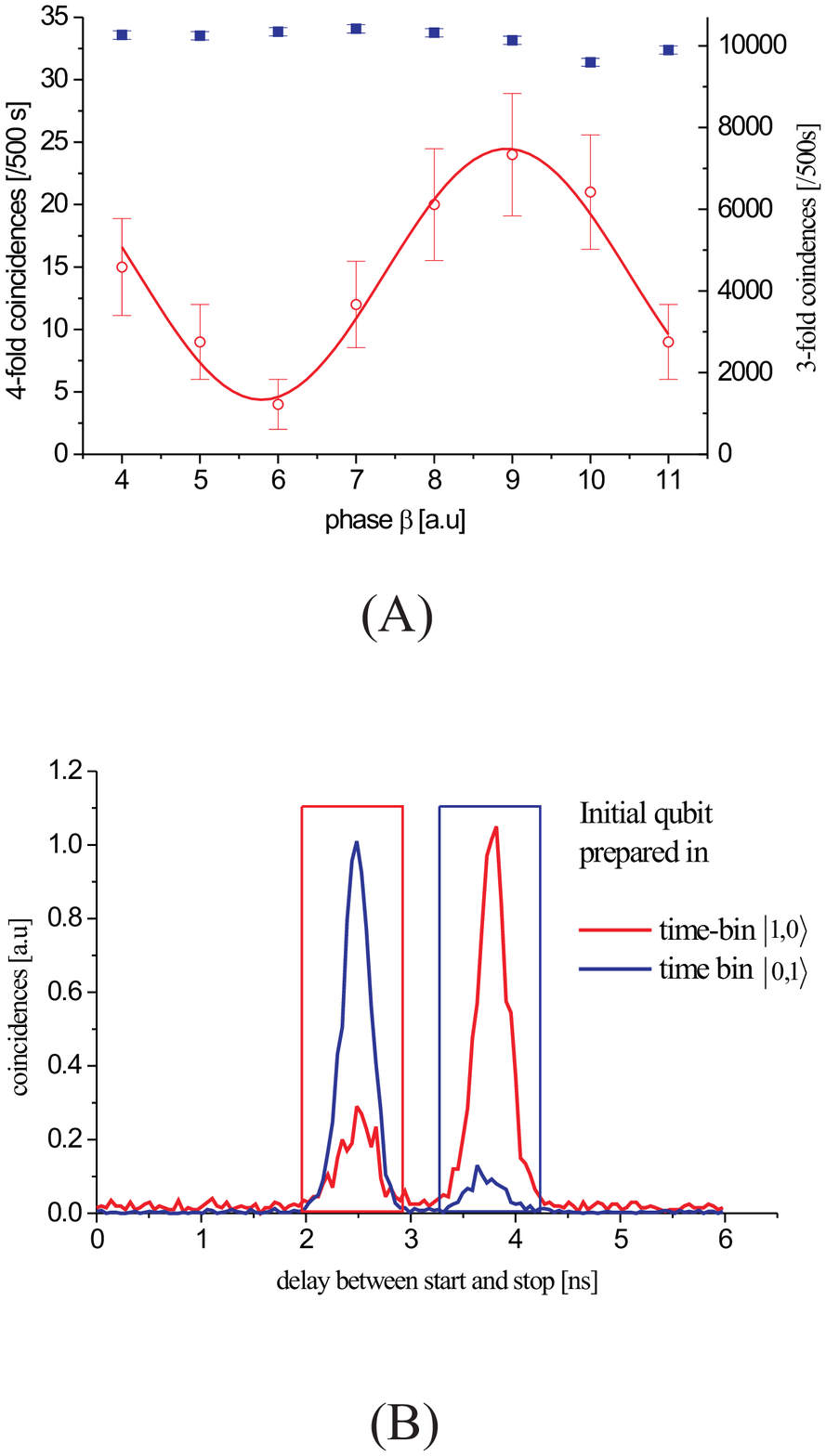}
\caption{{Experimental results: \textbf{a)}  Teleportation of a
qubit state constituted of a coherent superposition of two
time-bins (equatorial states). The red open circles represent
Bob's count rate conditioned on the projection on the
$\left|\Psi^-\right \rangle$ Bell state (four fold coincidences)
as a function of the phase $\beta$ in his interferometer. The
clear interference pattern with visibility $V=(70\pm5)\,\%$,
confirms that teleportation takes place, with a fidelity of $(85
\pm 2.5)\,\%$. The blue squares represent the three fold
coincidence count rate between laser pulse, one of Charlie's
detectors ($C_1$) and Bob's detector. They illustrate the
stability of the setup. \textbf{b)} Teleportation of qubits
$\left|1,0\right\rangle$ (north pole, red curve) and
$\left|0,1\right\rangle$ (south pole, blue curve). The
corresponding fidelities are $(77\pm3)\,\%$ and $(88\pm3)\,\%$
respectively. As indicated in Eq.\ref{4}, the projection onto the
$\left|\Psi^-\right\rangle$ Bell state leads to a bit flip of the
teleported state.}} \label{results}
\end{figure}

The preparation of the two other states, represented by the north
and south poles, corresponds to the variable coupler at settings
0\,\% and 100\,\%, respectively. That was realized by using two
different fibres of appropriate lengths. For the measurement, Bob
uses only one fibre and looks for detections at appropriate times.
As reported in Eq.\ref{4}, when Alice sends such a state, Bob
receives the orthogonal state, i.e. he should bit flip his qubit
to recover Alice's state. The corresponding fidelity is the
probability of detecting the right state when measuring in the
north-south basis,
$F_{poles}=\frac{R_{correct}}{R_{wrong}+R_{correct}}$ , see
Fig.4b. The measured fidelity for the $\left|1,0\right\rangle$
input state is $(77\pm 3)\,\%$ and for the
$\left|0,1\right\rangle$ input state $(88 \pm 3)\,\%$. Accordingly
the mean value is $F_{poles}= (82.5 \pm 3)\,\%$. The different
results for the two input states is due to the fact that in our
experimental set-up the detection of the first photon, in mode
$C_1$, triggers the two other detectors $C_2$ and $B$. When we
prepare the state $\left|1,0\right\rangle$ the first detection is
mostly due to Alice's photon since, as explained above, we produce
8 times more qubits to be teleported than entangled pairs. Hence,
the two detectors $C_2$ and $B$ are often triggered without any
photon present, leading to an increasing number of accidental
coincidences, i.e. wrong events. When preparing the other state
$\left|0,1\right\rangle$ the first detection in mode $C_1$ can
only be due to a photon coming from the EPR source or to a dark
count. Since these events occur much less frequently than in the
first mentioned case, the corresponding teleportation fidelity is
less affected by accidental coincidences.

From these results, we conclude that the overall mean fidelity is
$\bar F=(81.2 \pm 2.5)\,\%$ (Eq.\ref{7}). This value is six
standard deviations above the maximum fidelity of 66.7\,\%
achievable with the best protocol using no entanglement. Note that
the above reported result takes into account noise produced by
Bob's analyzer. Strictly speaking, this noise should not be
attributed to the teleportation scheme. Hence, the achieved
teleportation fidelity is actually higher, about
$\frac{2}{3}0.86+\frac{1}{3}0.83=0.85$. However, for practical
applications, the significant value is $\bar F$.

The difference between the experimental results and the ideal
theoretical case can be due to various imperfections: \textbf{(i)}
Our Bell state measurement is not perfect; although we tried to
detect only the events when there is only one photon in each mode,
there is still 11\,\% of chances that we make a spurious
coincidence. This value can be found in the noise of the
teleportation of the $\left|0,1\right\rangle$ input state, which
is essentially due to the creation of double entangled pairs.
Further reduction of the fidelity might be due to different
polarization or spectra of the two photons when arriving at the
beam splitter or to remaining temporal distinguishability.
\textbf{(ii)} The creation and analysis of the qubits is not
perfect, e.g, the two qubits entanglement leads to a fringe
visibility limited to 95\,\% \cite{marcikic02}. \textbf{(iii)}
Detector dark counts also reduce the measured fidelity. Finally,
note that the fact that \textbf{a)} the teleportation occurs only
when there is a projection onto the state $\left|
\Psi^{-}\right\rangle$ and that \textbf{b)} only one eighth of the
qubits sent by Alice are teleported renders our realization
probabilistic, even assuming perfect detectors. This is a drawback
from a fundamental point of view. However, if quantum
teleportation is used as quantum relay (see next chapter and
fig.\ref{relay}) in quantum cryptography, then the probabilistic
nature of our teleportation scheme will only affect the count rate
but not the quality of the quantum relay. For this application our
scheme is useful.

\section{Quantum relays}

\begin{figure}[h]
\includegraphics[width=8.43cm]{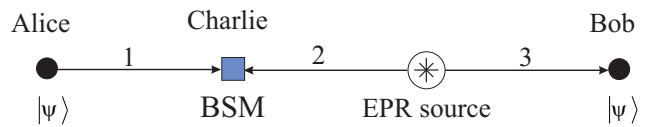}
\caption{{Quantum teleportation used as a quantum relay. The
quantum channel between Alice and Bob is divided in three
sections. BSM means Bell State Measurement. A conclusive BSM at
Charlie's insure that one photon carrying Alice's state
$\left|\Psi\right\rangle$ has left the EPR source towards Bob.}}
\label{relay}
\end{figure}

Presently, the only potential application of quantum teleportation
is as quantum repeaters for quantum cryptography \cite{gisin02}.
Actually, a fully developed quantum repeater would also require
quantum memories \cite{briegel98}, but we shall see that quantum
teleportation without quantum memory (so called quantum relay) can
already extend the range of quantum cryptography from tens of km
to hundreds, though not to unlimited ranges. The basic idea is as
follows \cite{gisin02,waks02,jakobs02}. In quantum cryptography,
the noise is dominated by the detector dark counts; hence the
noise is almost independent of the distance. The signal, however,
decreases exponentially with distance because of the attenuation.
With realistic numbers, this sets a limit close to 80 km. But, if
one could check at some points along the quantum channel whether
or not the photon is still there, one could refrain from opening
the detector when there is no photon. This simple idea is
unpractical because it requires (presently unrealistic) photon
number quantum non demolition measurements \cite{grangier98}.
However, consider a channel divided into sections, e.g. 3 sections
as shown in Fig.\ref{relay}. Assume that the photon sent by Alice
down the first section is teleported to Bob using the EPR photon
pair generated in-between sections 2 and 3. Two photons travel
towards Bob, but one towards Alice. But, since the time ordering
of the measurements is irrelevant \cite{jennewein02}, one may
consider that the logical qubit propagates all the way from Alice
to Bob. Accordingly, the Bell state measurement and the 2-photon
source act on this logical qubit as a non demolition measurement!

\section{Conclusion}

Quantum teleportation is a fascinating prediction of quantum
mechanics which shakens basic concepts like object, information,
space-time. We have reported the first long distance demonstration
of this protocol. Using 2\,km of standard telecom optical fibres,
we teleported a qubit carried by a photon of 1.3\,$\mu m$
wavelength to a qubit in another lab carried by a photon of
1.5\,$\mu m$ wavelength. The photon to be teleported and the
necessary entangled photon pairs were created in two different
non-linear crystals. The measured mean fidelity of 81.2\,\% is
sufficient to demonstrate the basic principle. However,
admittedly, work towards useful teleportation set-ups, e.g. as
quantum relays as explained above, still requires a lot of effort
and ideas, mainly to improve the stability of the experiment.


\begin{thebibliography}{99}

\bibitem{brassard93} C.H. Bennett, G. Brassard, C. Crépeau, R. Jozsa,
A. Peres, and W.K. Wootters, Teleporting an unknown quantum state
via dual classical and Einstein-Podolsky-Rosen channels, {\it
Phys. Rev. Lett.}  \textbf{70}, 1895-1899 (1993)

\bibitem{einstein35} A. Einstein, B. Podolsky, and N. Rosen,
Can Quantum-Mechanical Description of Physical Reality Be
Considered Complete?, {\it Phys. Rev.} \textbf{47}, 777-780 (1935)

\bibitem{gisin02} N. Gisin, G. Ribordy, W. Tittel, and H. Zbinden,
Quantum cryptography, {\it Reviews of Modern Physics} \textbf{74},
145-195 (2002)

\bibitem{gottesman99} D. Gottesman and I.L. Chuang,
Demonstrating the viability of universal quantum computation using
teleportation and single-qubit operations, {\it Nature}
\textbf{402}, 390-393 1999

\bibitem{briegel98} H.-J. Briegle, W. Dur, J.I. Cirac, and P. Zoller,
Quantum Repeaters: The role of Imperfect local Operations in
Quantum Communication, {\it Phys. Rev. Lett.} \textbf{81},
5932-5935 (1998)

\bibitem{boschi98} D. Boschi, S. Branca, F. De Martini, L. Hardy, and S. Popescu,
Experimental Realization of Teleporting an Unknown Pure Quantum
State via Dual Classical and Einstein-Podolsky-Rosen Channels,
{\it Phys. Rev. Lett.} \textbf{80}, 1121-1125 (1998)

\bibitem{bouwmeester97} D. Bouwmeester, J.W. Pan, K. Mattle, M. Eibl, H. Weinfurter, and A. Zeilinger,
Experimental quantum teleportation, {\it Nature} \textbf{390},
575-579 (1997)

\bibitem{furusawa98} A. Furusawa, J. L. Sørensen, S. L. Braunstein, C. A. Fuchs, H. J. Kimble, and E. S.
Polzik, Unconditional Quantum Teleportation, {\it Science}
\textbf{282}, 706-709 (1998)

\bibitem{nielsen98} Nielsen M.A, Knill E, and Laflamme R., Complete quantum teleportation using nuclear magnetic resonance, {\it Nature} \textbf{396}, 52-55 (1998)

\bibitem{kim01} Y.-H. Kim, S. P. Kulik, and Y. Shih, Quantum Teleportation of a Polarization State with a Complete Bell State Measurement, {\it Phys. Rev. Lett.} \textbf{86}, 1370-1373 (2001)

\bibitem{lombardi02} E. Lombardi, F. Sciarrino, S. Popescu, and F. De Martini, Teleportation of a Vacuum--One-Photon Qubit, {\it Phys. Rev. Lett.} \textbf{88}, 070402 (2002)

\bibitem{babichev02} S.A. Babichev, J. Ries, and A.I. Lvovsky, Quantum scissors: teleportation of single-mode optical states by mean of a nonlocal single photon, {\it Quant-ph/} 0208066 (2002)

\bibitem{bowen02} W.P. Bowen, N. Treps, B.C. Buchler, R. Schnabel, T.C. Ralph, H.-A. Bachor, T. Symul, P.K. Lam,
Experimental investigation of continuous variable quantum
teleportation, {\it Quant-ph/}0207179 (2002)

\bibitem{jennewein02} T. Jennewein, G. Weihs, J.-W. Pan, and A. Zeilinger, Experimental Nonlocality Proof of Quantum Teleportation
 and Entanglement Swapping, {\it Phys. Rev. Lett} \textbf{88} 017903 (2002)

\bibitem{zukowski93} M. Zukowski, A.Zeilinger, M. A. Horne, and A. K. Ekert
"Event-ready-detectors" Bell experiment via entanglement swapping,
{\it Phys. Rev. Lett.} \textbf{71}, 4287-4290 (1993)

\bibitem{pan98} J.-W. Pan, D. Bouwmeester, H. Weinfurter, and A. Zeilinger,
Experimental Entanglement Swapping: Entangling Photons That Never
Interacted , {\it Phys. Rev. Lett.} \textbf{80}, 3891-3894 (1998)

\bibitem{lutkenhaus99} N. L\"utkenhaus, J. Calsamiglia, and K.-A. Suominen,
Bell measurements for teleportation, {\it Phys. Rev. A}
\textbf{59}, 3295-3300 (1999)

\bibitem{braunstein98} S. L. Braunstein and H. J. Kimble,
Teleportation of Continuous Quantum Variables, {\it Phys. Rev.
Lett.} \textbf{80}, 869-872 (1998)

\bibitem{tittel01} W. Tittel and G. Weihs,
Photonic entanglement for fundamental tests and quantum
communication , {\it Quantum Information \& Computation}
\textbf{1}, 3 (2001)

\bibitem{brendel99} J. Brendel, W. Tittel, H. Zbinden, and N.Gisin,
Pulsed energy-time entangled twin-photon source for quantum
Communication, {\it Phys. Rev. Lett.} \textbf{82}, 2594-2597
(1999)

\bibitem{thew02} R. T. Thew, S. Tanzilli, W. Tittel, H. Zbinden, and N. Gisin,
Experimental investigation of the robustness of partially
entangled photons over 11km,  {\it Phys. Rev. A} \textbf{66},
062304 (2002)

\bibitem{tittel00} W. Tittel, J. Brendel, H. Zbinden, and N. Gisin, Quantum Cryptography using entangled photons in energy-time Bell states, {\it Phys. Rev. Lett.} \textbf{84}, 4737-4740 (2000)

\bibitem{riedmatten02} H. De Riedmatten, I. Marcikic, H. Zbinden, and N. Gisin
Creating high dimensional entanglement using mode-locked laser,
{\it Quant. Inf. Comput} \textbf{2}, 425-433 (2002)

\bibitem{zbinden01} see for instance H. Zbinden, J. Brendel, N. Gisin, and W. Tittel,
Experimental test of nonlocal quantum correlation in relativistic
configurations, {\it Phys. Rev. A} \textbf{63}, 022111 (2001) and
references therein


\bibitem{noe99} R. Noé, D. Sandel, M. Yoshida-Dierolf, S. Hinz, V. Mirvoda, A. Schöpflin, C. Glingener, E. Gottwald, C. Scheerer, G. Fischer, T. Weyrauch, and W. Haase, Polarization
mode dispersion compensation at 10, 20, and 40 Gb/s with various
optical equalizers, {\it IEEE J. lightwave Tech.} \textbf{17},
1602-1617 (1999)

\bibitem{michler96} M. Michler, K.Mattle, H. Weinfurter, and A. Zeilinger, Interferometric Bell State Analysis, {\it Phys.Rev. A} \textbf{53}, 1209 (1996)

\bibitem{marcikic02} I.Marcikic, H. de Riedmatten, W.Tittel, H.Zbinden, and N.Gisin,
Femtosecond Time-Bin Entangled Qubits for Quantum Communication,
{\it Phys. Rev. A} \textbf{66}, 062308 (2002)

\bibitem{riedmattendip02} H. de Riedmatten, I.Marcikic, W.Tittel, H.Zbinden, and N.Gisin, Quantum interferences with photon pairs created in spatially separated sources, {\it
quant-ph/}0208174 (in press {\it Phys.Rev. A})

\bibitem{owens94} P.C.M. Owens, J.G. Rarity, P.R. Tapster, D. Knight, and P.D. Townsend
Photon counting with passively quenched germanium avalanche, {\it
Appl. Opt.} \textbf{33}, 6895-6901 (1994)

\bibitem{ribordy98} G. Ribordy, J.D. Gautier, H. Zbinden, and N. Gisin,
Performance of InGaAs/InP avalanche photodiodes as gated-mode
photon counters, {\it Appl. Opt.} \textbf{37} (12), 2272-2277
(1998)

\bibitem{stucki01} D.Stucki {\it et al.}, 
Photon counting for quantum key distribution with Peltier cooled
InGaAs/InP APD's,  {\it J. of Mod. Optics} \textbf{48} (13), 1967
(2001)

\bibitem{linares01} A. Lamas-Linares {\it et al.},
Stimulated emission of polarization-entangled photons, {\it
Nature} \textbf{412}, 887-890 (2001)

\bibitem{massar95} S. Massar and S. Popescu,
Optimal exctraction of information from finite quantum ensambles,
{\it Phys. Rev. Lett.} \textbf{74}, 1259-1263 (1995)

\bibitem{waks02} E. Waks {\it et al.},
Security of Quantum Key Distribution with Entangled Photons
Against Individual Attacks, {\it Phys. Rev. A} \textbf{65}, 052310
(2002)

\bibitem{jakobs02} B. C. Jacobs {\it et al.}, Quantum relays and noise suppression using linear optics, {\it Phys. Rev.
A} \textbf{66}, 052307 (2002)

\bibitem{grangier98} P. Grangier {\it et al.}, Quantum non-demolition measurements in optics, {\it Nature}
\textbf{396}, 537-542 (1998)


\end{thebibliography}
\end{document}